\documentclass[letterpaper, 10pt, conference]{ieeeconf}      

\IEEEoverridecommandlockouts                              

\usepackage{graphicx} 
\usepackage{amsmath} 
\usepackage{amssymb}  

\usepackage{amsthm}
\usepackage{pgfplots}
\usepackage{subfig}
\usepackage{multirow}

\usetikzlibrary{external}

\newtheoremstyle{algorithm}
{\topsep}
{\topsep}
{}
{}
{\scshape}
{.}
{ }
{\thmname{#1}\thmnumber{ #2}\thmnote{ (#3)}}
\theoremstyle{algorithm}
\newtheorem{algorithm}{Algorithm}

\title{\LARGE \bf
  Lasso Regularization Paths for NARMAX Models\\ via Coordinate Descent*
}

\author{Ant\^{o}nio H. Ribeiro$^{1}$ and Luis A. Aguirre$^{2}$
\thanks{*This work has been supported by the Brazilian agencies CAPES, CNPq
    and FAPEMIG.}
\thanks{$^{1}$Graduate  Program in  Electrical  Engineering,  Federal
  University  of  Minas Gerais - Av. Ant\^{o}nio Carlos 6627, 31270-901,
  Belo Horizonte. {\tt\small antonio-ribeiro@ufmg.br}}%
\thanks{$^{2}$Department of Electronic Engineering,  Federal  University
  of  Minas Gerais. {\tt\small aguirre@ufmg.br}}%
}
 
\begin{document}

\maketitle
\thispagestyle{empty}
\pagestyle{empty}

\begin{abstract}
  We propose a new algorithm  for estimating \hbox{NARMAX} models with $L_1$ regularization for models represented as a linear combination of basis functions. Due to the $L_1$-norm penalty the Lasso estimation tends to produce some coefficients that are exactly zero and hence gives interpretable models. The novelty of the contribution is the inclusion of error regressors in the Lasso estimation (which yields a nonlinear regression problem).  The proposed algorithm uses cyclical coordinate descent to compute the parameters of the NARMAX models for the entire regularization path. It deals with the error terms by updating  the regressor matrix along with the parameter vector.  In comparative timings we find that the modification does not reduce the computational efficiency of the original algorithm and can provide the most important regressors in very few inexpensive iterations. The method is illustrated for linear and polynomial models by means of two examples.
\end{abstract}

\section{Introduction}
\label{sec:introduction}

The \textit{Lasso}
(\textit{least absolute shrinkage and selection operator})~\cite{tibshirani_regression_1996}
proposed in 1996 by Robert Tibshirani
is a popular method for regularizing least squares regression using $L_1$
penalization to achieve sparse solutions. Like subset selection
methods (e.g. forward-stepwise regression~\cite{chen_orthogonal_1989,
  billings_identification_1989})
it allows the data analyst to control the model complexity
in order to avoid overfitting the data. In the case of subset
selection methods, the model complexity can be restricted
by limiting the number of regressors to enter the model.
For the Lasso, the regularization weight
can be used to control the degrees of freedom of the model.
Lasso has the advantage of being more \textit{stable} than subset selection.
Here \textit{stable} is used in the sense defined in~\cite{breiman_heuristics_1996}:
small modifications in the training data do not cause large changes in the optimal
choice of model complexity.

The application of Lasso and variations (e.g. grouped Lasso~\cite{meier_group_2008} and
\textit{elastic net}~\cite{zou_regularization_2005}) for building models of
dynamic systems and time series have received considerable attention
in the last years. They have been used for the identification of
nonparametric~\cite{chiuso_nonparametric_2010,  calafiore_leading_2017},
polynomial~\cite{kukreja_least_2006} and posynomial~\cite{calafiore_sparse_2015}
dynamic models.
In~\cite{nardi_autoregressive_2011,  basu_regularized_2015}
its application to autoregressive time series was studied, with the estimator properties
derived. A recursive online version of Lasso was proposed
in~\cite{cantelmo_adaptive_2010}. The combination of Lasso
with a pruning algorithm for structure selection was studied in~\cite{bonin_narx_2010}.
And, in~\cite{ha_model_2015} Lasso was used in an instrumental variable setup
for selecting the order of linear continuous models.

In this paper we study the application of Lasso regularization
for the estimation of
NARMAX (\textit{Nonlinear Autoregressive Moving Average With Exogenous Input})
models. NARMAX models include an \textit{error model} that is estimated
together with the \textit{process model}. Hence,
NARMAX models may yield consistent results even in the presence of colored
equation errors. However, the presence of noise terms also results
in a nonlinear regression which requires the solution of a non-convex optimization
problem.

In the original paper~\cite{tibshirani_regression_1996},
the Lasso solution was obtained by solving quadratic programming
problems. This approach, however, did not scale very well and was not very
transparent~\cite{tibshirani_regression_2011}.  The LARS (Least Angle Regression)
algorithm~\cite{efron_least_2004} proposed in 2004
solves the entire regularization path with a similar computational
cost to the least squares algorithm.
A competing approach that has been proved the most efficient
(according to benchmarks presented in~\cite{friedman_regularization_2010})
is to use coordinate descent optimization to find the Lasso path,
by solving ``one-at-a-time'' unidimensional optimization problems
along the coordinates. This idea has been proposed very
early~\cite{fu_penalized_1998} but its potential
was only fully appreciated later,
after studies and efficient implementations~\cite{friedman_pathwise_2007,
  friedman_glmnet:_2009, friedman_regularization_2010}
demonstrated its great potential.

The algorithm proposed here is based
on the coordinate descent algorithm~\cite{friedman_regularization_2010}
and assume the NARMAX model is given by a linear combination
of basis function. It uses the problem structure in order to solve the sequence of
non-convex non-differentiable problems efficiently.
It deals with the non-linearities of NARMAX estimation by updating the error
model along with the solution, what may render some computational tricks
proposed by~\cite{friedman_pathwise_2007}
impossible. Nevertheless the implementation is efficient and
applicable to a large range of problems.

It is important to acknowledge~\cite{wang_regression_2007, yoon_penalized_2013}
for also including noise terms in the Lasso regression problem. Our formulation, however,
consider an \textit{autoregressive with exogenous input} process model and
a \textit{moving average} error model.
Theirs formulation, on the other hand, considered a \textit{finite impulse response (FIR)}
process model and an
\textit{autoregressive} error model.

The rest of the paper is organized as follows: Section~\ref{sec:lasso-pathw-coord}
and~\ref{sec:pred-error-meth} provide the required background on, respectively,
Lasso and NARMAX models. The proposed algorithm
is described in Section~\ref{sec:coord-optim-algor}. Test results and implementation details are described in
Section~\ref{sec:impl-test-results} and final comments are provided in
Section~\ref{sec:final-comments}.

\section{Lasso and the Pathwise Coordinate Optimization Algorithm}
\label{sec:lasso-pathw-coord}

\subsection{Lasso}

Consider the usual setup for linear regression, being
${\mathbf{X}\in \mathbb{R}^{N\times p}}$
a matrix containing observations of independent variables
and ${\mathbf{y}\in \mathbb{R}^{N}}$ a vector
containing the corresponding dependent variable.
We assume that all variables
have been centered and have zero mean.

The Lasso solution of this regression problem
is given by the solution of the following minimization
problem:
\begin{equation}
  \label{eq:lasso_problem}
  \min_{\boldsymbol{\theta}} \tfrac{1}{2}\|\mathbf{y} - \mathbf{X} \boldsymbol{\theta}\|_2^2 +
  \lambda\|\boldsymbol{\theta}\|_1,
\end{equation}
where $\boldsymbol{\theta}\in \mathbb{R}^{p}$ is a vector containing
parameters that we wish to estimate from observation data and
$\lambda$ weighs the regularization term.

This formulation produces sparse solutions due to the
penalty term introducing non-differentiable corners along the regions where $\theta_i = 0$.
The number of non-zero terms
depends on the value of $\lambda$. The larger the value of
$\lambda$, the lesser degrees of freedom are given to the solution.

The Lasso minimization problem can be
interpreted as the Lagrangian formulation of a least-squares problem subject
to the constraint $\|\boldsymbol{\theta}\|_1 \le \Delta$, for $\Delta$ dependent on $\lambda$.
Alternatively, it can be viewed
as the maximum a posteriori parameter estimation
considering a Laplacian prior. Refer to~\cite{hastie_statistical_2015}
for a complete discussion on the method.

\subsection{Pathwise Coordinate Optimization}

Consider a coordinate descent step for solving~(\ref{eq:lasso_problem}).
Be $\theta_j$ the $j$-th component of the parameter vector,
suppose that all the components $\theta_i$ for $i \not= j$
are fixed and we want to optimize~(\ref{eq:lasso_problem})
with respect to $\theta_j$. Simple manipulations
show that this yields the unidimesional optimization problem:
\begin{equation}
  \label{eq:unidimensional_lasso}
  \min_{\theta_j} \tfrac{1}{2}\|\mathbf{x}_j \|_2^2\theta_j^2 -
  (\mathbf{y} - \sum_{i\not= j}\mathbf{x}_i \theta_i)^T\mathbf{x}_j \theta_j +
  \lambda|\theta_j| + C,
\end{equation}
where $\mathbf{x}_i$ is the $i$-th column of $\mathbf{X}$;
and, ${C = \tfrac{1}{2}\|\mathbf{y} - \sum_{i\not= j}\mathbf{x}_i \theta_i\|^2
+ \lambda \sum_{i\not= j}|\theta_i|}$ is the term containing all fixed components.

The analytic solution of~(\ref{eq:unidimensional_lasso}) can be found
by minimizing the corresponding polynomial of degree 2 for three different
situations: $ \theta_i>0$; $\theta_i=0$; and, $\theta_i<0$. This
yields the optimal coordinate update:
\begin{equation}
  \label{eq:coordinate_wise_update}
  \theta_j \leftarrow \frac{1}{\|\mathbf{x}_j\|^2}S\Big(
    (\mathbf{y} - \sum_{i\not= j}\mathbf{x}_i \theta_i)^T\mathbf{x}_j;~
    \lambda\Big),
\end{equation}
where $S(z, \lambda)$ stands for the soft-thresholding
operator:
\begin{equation}
  \label{eq:soft_threshoulding_operator}
  S(z;~ \lambda) =
  \begin{cases}
    z - \lambda & \text{if } z > \lambda\\
    0 & \text{if } -\lambda < z < \lambda\\
    z + \lambda & \text{if } z < -\lambda.
  \end{cases}
\end{equation}
Thus the algorithm applies the update~(\ref{eq:coordinate_wise_update})
cyclically along the coordinates until the solution converges.
Conditions for convergence are given in~\cite{tseng_convergence_2001}.

\subsection{Computing the Update}
\label{sec:computing-update}

In~\cite{friedman_regularization_2010} two different ways of storing
and updating the computation of
$\tilde{\mathtt{r}}_{j} = (\mathbf{y} - \sum_{i\not= j}\mathbf{x}_i \theta_i)^T\mathbf{x}_j$ are discussed.

The so-called \textit{naive update} approach
keeps an updated value of the residual vector
$\mathbf{r} = \mathbf{y} - \sum_{i=1}^p\mathbf{x}_i \theta_i$ stored
and computes
$\tilde{\mathtt{r}}_{j} \leftarrow (\mathbf{r} + \mathbf{x}_j \theta_j)^T\mathbf{x}_j$.
This yields a computational cost of $\mathcal{O}(N)$ per iteration
and $\mathcal{O}(p\cdot N)$ for each complete cycle
through all $p$ variables.

The \textit{covariance update} keeps values of $\mathbf{y}^T\mathbf{x}_j$
and $\mathbf{x}_i^T\mathbf{x}_j$ stored, and computes
$\tilde{\mathtt{r}}_{j} \leftarrow (\mathbf{y}^T\mathbf{x}_j -
\sum_{i\not= j}\mathbf{x}_i^T\mathbf{x}_j \theta_i)$. That way,
each time a new variable enters the model there is an
associated computational cost of $\mathcal{O}(p\cdot N)$
due to the computation of all dot products $\mathbf{x}_i^T\mathbf{x}_j$.
For the remaining iterations, however, the cost of the iteration is $\mathcal{O}(p\cdot m)$,
where $m<p$ is the number of non-zero variables.
Hence, for the covariance approach, $\mathcal{O}(N)$
computations are not required at each steps.

The adapted version of coordinate descent optimization for NARMAX models
described in Section~\ref{sec:coord-optim-algor} uses the \textit{naive update}
because the proposed modifications renders the use of the more
efficient \textit{covariance update} impossible.

\subsection{Warm Start}

Reference \cite{friedman_pathwise_2007} points out the role of
\textit{warm starts} in the efficient computation of the entire
regularization path. The procedure consists of, starting
with $\lambda = \lambda_{\max}$, computing the solution
for a decreasing sequence of values of $\lambda$,
using the estimated parameter vector at the last iteration
as initial guess to be refined for the current value of lambda.

Here $\lambda_{\max}$ denotes the smallest value of $\lambda$
for which the entire parameter vector $\boldsymbol{\theta}$ is zero.
A minimum value ${\lambda_{\min} < \lambda_{\max}}$ is selected
and a decreasing sequence of $K$ values (in log-scale)
between $\lambda_{\max}$ and $\lambda_{\min}$ is constructed.

It follows from~(\ref{eq:coordinate_wise_update}) that if
$\lambda >|\mathbf{y}^T\mathbf{x}_j|$ all the updates are going to be zero.
Hence, we can set $\lambda_{\max} = \max_j  |\mathbf{y}^T\mathbf{x}_j|$.
Typical values of $K$ and $\lambda_{\min}$ 
are, according to~\cite{friedman_regularization_2010}, ${K = 100}$ and
$\lambda_{\min} = 0.001\lambda_{\max}$.

\subsection{Active Set Convergence}
\label{sec:active-set-conv}

Some speedup can be obtained by, instead of cycling along all the $p$
variables every time, to organize the iterations around the \textit{active set}
(non-zero variables)~\cite{friedman_regularization_2010}. That is,
after a complete cycle through all the variables, we iterate only on the active
set until convergence. A new complete cycle through the complete set
follows, interrupting the processes if no change on active set is found.

\section{NARMAX Models}
\label{sec:pred-error-meth}

Consider the data set
$\mathcal{Z} = \{(\mathbf{u}[k], y[k]), k = 1, 2, \hdots, N\},$
containing  a sequence of sampled inputs-output pairs.
Here ${\mathbf{u}[k]\in \mathbb{R}^{N_u}}$ 
is a vector containing all the inputs
and ${y[k]\in \mathbb{R}}$ is the scalar output.
The output $y[k]$ is correlated with its own past
values and with past input values.
The focus of this paper is trying to find a difference equation model
that best describe the observed data.

\subsection{Optimal Predictor}

To study the previously described problem it is assumed that for a given input
sequence ${\mathbf{u}[k]}$ the output was generated by a \textit{``true system''},
described by the following difference equation:
\begin{footnotesize}
  \begin{multline}
    \label{eq:narmax_assumption}
    y[k] = F(y[k-1], \hdots, y[k-n_y],
             \mathbf{u}[k-\tau_d], \hdots, \mathbf{u}[k-n_u], \\
             v[k-1], \hdots, v[k-n_v];~\boldsymbol{\theta}^*) + v[k],
  \end{multline}
\end{footnotesize}
where $F$ and $\boldsymbol{\theta}^*$ are the \textit{``true''} function
and parameter vector describing the
system, $n_y$, $n_u$ and $n_v$ are the maximum input, output
and error lag and $\tau_d$ is the input-output delay. The
assumption that a finite number of past terms
can be used to describe the output is implicit in this model.

Furthermore, $v[k]\in \mathbb{R}$ is a random variable
that causes the deviation of the deterministic model from its true value.
We assume here that $v[k]$ is a white random process (which implies
it has zero mean and that $v[k]$ is uncorrelated with $v[l]$
for $l \not= k$). The capability of the above
model to represent colored noise comes from
the presence of lagged error terms $v[k-i]$ in function
$F$ arguments.

The simplified notation $\underline{\mathbf{y}}_{[k-1]}$,
$\underline{\mathbf{v}}_{[k-1]}$ and
$\underline{\mathbf{u}}_{[k-1]}$
will be used to represent
the vectors
${\footnotesize \big[y[k-1], \hdots,
  y[k-n_y]\big]^T}$, ${\footnotesize \big[v[k-1], \hdots,
  v[k-n_v]\big]^T}$ and
${\footnotesize \big[\mathbf{u}[k-\tau_d]^T, \hdots,
  \mathbf{u}[k-n_u]^T\big]^T}$. Using this new notation,
equation~(\ref{eq:narmax_assumption}) could be compactly rewritten as:
$y[k] = F(\underline{\mathbf{y}}_{[k-1]},
\underline{\mathbf{u}}_{[k-1]},
\underline{\mathbf{v}}_{[k-1]};~\boldsymbol{\theta}^*)+v[k]$.

If the measured values of $y$ and $\mathbf{u}$ are known at all instants
previous to $k$, the optimal prediction
of $\mathbf{y}[k]$ is the following conditional expectation:
\footnote{The expectation is the optimal prediction
  in the sense that the expected squared prediction error is
  minimized.}
\begin{equation}
  \label{eq:conditional_expectation}
  \hat{y}_*[k] = E\left\{y[k]~\Big\vert~
  \underline{\mathbf{y}}_{[k-1]}, 
  \underline{\mathbf{u}}_{[k-1]}\right\},
\end{equation}
where the notation $\hat{y}_*[k]$ is used to denote the optimal prediction.
Since $v[k]$ is a white process with zero mean, it follows that:
\begin{equation}
  \hat{y}_*[k] =  F(\underline{\mathbf{y}}_{[k-1]},
                   \underline{\mathbf{u}}_{[k-1]},
                   \underline{\mathbf{v}}_{[k-1]};~\boldsymbol{\theta}^*).\nonumber 
\end{equation}
Hence, $v[k] = y[k] - \hat{y}_*[k]$,
and the optimal predictor can be defined as follows:
\begin{equation}
  \label{eq:optimal_predictor}
  \hat{y}_*[k] = F(\underline{\mathbf{y}}_{[k-1]},
                   \underline{\mathbf{u}}_{[k-1]},
                   \underline{\mathbf{y}}_{[k-1]} - {\underline{\hat{\mathbf{y}}}_*}_{[k-1]};
                   ~\boldsymbol{\theta}^*). 
\end{equation}

\subsection{Parameter Estimation and Linear-in-the-Parameters Functions}

The parameter vector $\boldsymbol{\theta}$ of a NARMAX model
can be estimated by solving:
\begin{equation}
  \label{eq:predictor_error_minimization}
  \min_{\mathbf{\theta}} \sum_{k=1}^N (y[k] - \hat{y}[k])^2.
\end{equation}
where $y[k]$ is the measured output value and the prediction
$\hat{y}[k]$ is defined similarly to~(\ref{eq:optimal_predictor}):
\begin{equation}
  \label{eq:optimal_predictor_est}
  \hat{y}[k] = F(\underline{\mathbf{y}}_{[k-1]},
                   \underline{\mathbf{u}}_{[k-1]},
                   \underline{\mathbf{y}}_{[k-1]} -
                   {\underline{\hat{\mathbf{y}}}}_{[k-1]};
                   ~\boldsymbol{\theta}). 
\end{equation}
This problem is non-convex and cannot be written as an ordinary least-squares
problem due to the recurrent definition of $\hat{y}[k]$.
For linear-in-the-parameters representations,
\textit{extended least squares} algorithm~\cite{chen_orthogonal_1989} can be used for finding the solution.

Consider that $F$ can be written
according to a basis expansion:
\begin{equation}
  \label{eq:basis_expansion}
  F(\mathbf{y}_{[k-1]}, \underline{\mathbf{u}}_{[k-1]}, \underline{\mathbf{e}}_{[k-1]})
  = \sum_{i=1}^p \theta_i\cdot x_i(\mathbf{y}_{[k-1]}, \underline{\mathbf{u}}_{[k-1]},
  \underline{\mathbf{e}}_{[k-1]}),
\end{equation}
where the variable $e$ is being used to indicate the difference
$e[k] = y[k] - \hat{y}[k]$ and $x_i(\cdot)$ is a linear or nonlinear
transformation (e.g. ${x_i = y[k-5]}$, ${x_i = y^2[k-1]u_1[k-2]e[k-2]}$
or ${x_i = \tanh(y[k-1])}$).

It follows that the minimization problem~(\ref{eq:predictor_error_minimization})
can be rewritten as:
\begin{equation*}
  \min_{\mathbf{\theta}} \sum_{k=1}^N \left(y[k] -
  \sum_{i=1}^p \theta_i\cdot x_i(\mathbf{y}_{[k-1]}, \underline{\mathbf{u}}_{[k-1]},
  \underline{\mathbf{e}}_{[k-1]})\right)^2,
\end{equation*}
or, in matricial form:
\begin{equation}
  \label{eq:matricial_formulation}
  \min_{\mathbf{\theta}} \|\mathbf{y} - \mathbf{X}_{(\mathbf{y}, \mathbf{u}, \mathbf{e})}
  \boldsymbol{\theta}\|^2,
\end{equation}
for which $\mathbf{y}\in \mathbb{R}^N$ is a vector containing $y[k]$ as its elements;
and, $\mathbf{X}_{(\mathbf{y}, \mathbf{u}, \mathbf{e})}\in \mathbb{R}^{N\times p}$ is a matrix with
the elements $x_i(\mathbf{y}_{[k-1]}, \underline{\mathbf{u}}_{[k-1]},
\underline{\mathbf{e}}_{[k-1]})$ organized such that
the index $k$ grows along the matrix rows and the index $i$ along
the matrix columns.

Extended least squares
(Algorithm~\ref{alg:extended_least_squares})
estimate the parameters by consecutively
solving linear least squares problems, approximating
$\mathbf{e}$ by the current residual vector
$\mathbf{r}$. A similar approach  is adopted in the algorithm 
proposed in the next section.

\begin{algorithm}[Extended Least Squares]
  \label{alg:extended_least_squares}
  Given an initial guess for the residual vector $\mathbf{r}^{(0)}$ and $i =0$; Repeat until
  ${\|\theta^{i} - \theta^{i-1}\|_{\infty} < \text{tolerance}}$.
  \begin{enumerate}
  \item
    Compute the matrix $X_{(\mathbf{y}, \mathbf{u}, \mathbf{r}^{(i)})}$.
  \item
    Find $\boldsymbol{\theta}^{i+1}$ that minimizes
    ${\|\mathbf{y} - \mathbf{X}_{(\mathbf{y}, \mathbf{u}, \mathbf{r}^{(i)})}
      \boldsymbol{\theta}\|^2}$.
  \item
    Update
    $\mathbf{r}^{(i+1)} \leftarrow \mathbf{y}[k] -
    X_{(\mathbf{y}, \mathbf{u}, \mathbf{r}^{(i)})}\boldsymbol{\theta}^{i+1}$.
  \item
    Set $i \leftarrow i+1$.
  \end{enumerate}
\end{algorithm}

\section{Coordinate Optimization Algorithm Applied to NARMAX models}
\label{sec:coord-optim-algor}

The Lasso regression problem that arises when estimating
NARMAX models (for basis expansion representations)
requires the solution of the following minimization problem:
\begin{equation}
  \label{eq:matricial_formulation_lasso_narmax}
  \min_{\mathbf{\theta}} \|\mathbf{y} - \mathbf{X}_{(\mathbf{y}, \mathbf{u}, \mathbf{e})}
  \boldsymbol{\theta}\|^2_2 +  \lambda\|\boldsymbol{\theta}\|_1.
\end{equation}
A coordinate descent algorithm to solve the above minimization problem for a decreasing
sequence of $\lambda$ is presented in Algorithm~\ref{alg:coordinate_optimization}.
This algorithm is very similar to what was described in Section~\ref{sec:lasso-pathw-coord}
for linear problems. It considers, however, that the matrix
$\mathbf{X}_{(\mathbf{y}, \mathbf{u}, \mathbf{e})}$ varies along the iterations.

This algorithm keeps stored an updated version of the residual $\mathbf{r}$
and  updates the matrix
$\mathbf{X}_{(\mathbf{y}, \mathbf{u}, \mathbf{e})}$
by approximating $\mathbf{e}$ by the current estimate of $\mathbf{r}$.

As mentioned in Section~\ref{sec:computing-update}, the more efficient \textit{covariance
  update} is not applicable for NARMAX models because the matrix
$\mathbf{X}_{(\mathbf{y}, \mathbf{u}, \mathbf{e})}$ changes along the iterations.
Hence, Algorithm~\ref{alg:coordinate_optimization} uses the \textit{naive update}
approach.

Besides using the \textit{naive update} approach, the only modification
that was made in order to cope with NARMAX models
is the update of the $j$-th column of $\mathbf{X}_{(\mathbf{y}, \mathbf{u}, \mathbf{e})}$
every iteration (step 1-\textit{a}). Assuming that the update of
each element of the matrix $\mathbf{X}_{(\mathbf{y}, \mathbf{u}, \mathbf{e})}$
has a computational cost of $\mathcal{O}(1)$,
the cost of updating an entire column is $\mathcal{O}(N)$ and therefore the asymptotic
cost of each iteration is not altered by this step.

\begin{algorithm}[NARMAX Coordinate Optimization]
  \label{alg:coordinate_optimization}
  Set  $\mathbf{r} \leftarrow \mathbf{y}$, $\boldsymbol{\theta} \leftarrow \mathbf{0}$
  and $\lambda \leftarrow \lambda_{\max}$.
  \begin{enumerate}
  \item
    Compute the solution of~(\ref{eq:matricial_formulation_lasso_narmax})
    for the given value of $\lambda$ by repeating the following steps
    until a convergence criterion is met
    $\left(\text{e.g. }\|\theta^{+} - \theta\|_{\infty} < \text{tolerance}\right)$:
    \begin{enumerate}
    \item
      Update the $j$-th column of the matrix $X_{(\mathbf{y}, \mathbf{u}, \mathbf{e})}$
      considering $\mathbf{e}$ equals to the current estimate of the residual
      $\mathbf{r}$. Call this column vector $\mathbf{x}_j$
    \item
      Find the next value of $\theta_j$ according to:
      \begin{equation}
        \label{eq:update_formula_narmax}
        \theta_j^+ \leftarrow \frac{1}{\|\mathbf{x}_j\|^2}S\Big(
        (\mathbf{r} + \mathbf{x}_j \theta_j)^T\mathbf{x}_j;~
        \lambda\Big).
      \end{equation}
    \item
      Update the residual:
      \begin{equation}
        \label{eq:update_residual_formula_narmax}
        \mathbf{r} \leftarrow \mathbf{r} -\mathbf{x}_j(\theta_j^+ - \theta_j).
      \end{equation}
    \item
      Update the parameter $\theta_j \leftarrow \theta_j^+$.
    \item
      Update the index $j$. As discussed in
      Section~\ref{sec:active-set-conv}, this update can be done such that $j$ circles
      through all the $p$ variables on a first step and, after
      that, iterates on the active set until convergence.
    \end{enumerate}
  \item
    Store the estimated parameter vector and decrease the value of lambda $\lambda$.
    Keep the values of $\boldsymbol{\theta}$, $\mathbf{r}$ and
    $X_{(\mathbf{y}, \mathbf{u}, \mathbf{e})}$ to be used as warm start for the next iteration.
  \end{enumerate}
\end{algorithm}

Practical aspects of this implementation are discussed in the next
session.

\section{Implementation and Test Results}
\label{sec:impl-test-results}

Next we present numerical examples illustrating the method.
In these examples we focus exclusively on linear and polynomial representations.
The algorithm  was implemented in Julia
and the code to run the examples is available in the GitHub repository:
\hbox{\textit{https://github.com/antonior92/NarmaxLasso.jl}}.

\subsection{Example 1: Linear Model}

The following linear system:
\begin{equation}
  \label{eq:first_order_linear_system}
  y[k] = 0.5 y[k-1] - 0.5 u[k-1] + 0.5 v[k-1] + v[k],
\end{equation}
is simulated for a sequence of randomly generated
inputs $u$ and null initial conditions. The values are drawn from a
standard Gaussian distribution and each generated
value is held for 5 samples. And, $v$ is
a white Gaussian process with standard deviation
$\sigma_v = 0.3$.

A window of $2000$ samples is used for training
and a different realization with $1000$ samples is used
to validate the model.

\begin{figure}[ht]
    \input{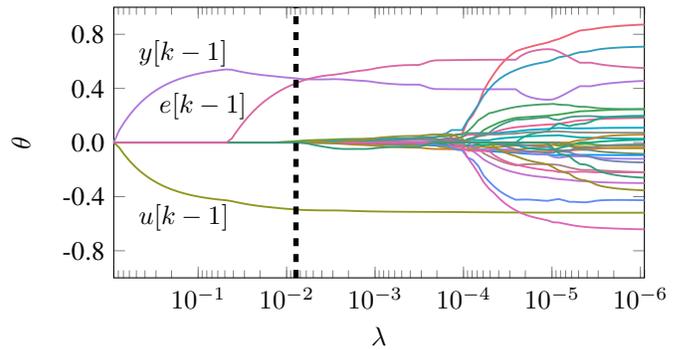}
    \caption{\textbf{(Example 1)} Estimated parameter vector $\boldsymbol{\theta}$
      as a function of the regularization parameter $\lambda$ (in log-scale).
      Parameters corresponding to $y[k-1]$, $u[k-1]$ and $e[k-1]$
      are annotated in the figure. A dashed vertical line
      indicates the value of $\lambda$ which yields the best validation results.}
  \label{fig:armax_example}
\end{figure}

We try to fit the following linear model to the training data:
\begin{equation}
  \label{eq:armax_model_initial}
  y[k] = \sum_{i=1}^{10}\theta_i y[k-i] + \sum_{i=1}^{10}\theta_{(i+10)} u[k-i]
         + \sum_{i=1}^{10}\theta_{(i+20)} e[k-i].
\end{equation}
Algorithm~\ref{alg:coordinate_optimization}
is used to find the parameter
vector that minimizes (\ref{eq:matricial_formulation_lasso_narmax})
for a sequence of decreasing values of $\lambda$.
The result is presented in Figure~\ref{fig:armax_example}
which shows the estimated parameters as a function of
$\lambda$. Notice that the procedure generates parameters that are
exactly zero. For high values of $\lambda$ most of the
parameters are zero and, as we decrease it, more and more terms
are included.
The values $y[k-1]$, $u[k-1]$ and $e[k-1]$ are the first terms to enter
the model and, as $\lambda$ approaches zero, other terms enter in the model as well.

We simulate each of the obtained models using the \textit{validation window}
and select the value of $\lambda$ that yields the smallest sum of absolute
errors between the estimated model free-run simulation and the observed
values. This value of $\lambda$ is indicated by a dashed vertical line in the figure.
For this value of $\lambda$ the estimated model is:
\begin{equation*}
  y[k] = 0.48 y[k-1] -   0.50 u[k-1]
         + 0.44 e[k-1].
\end{equation*}

\subsection{Example 2: Nonlinear Model}

Consider the non-linear system:~\cite{chen_non-linear_1990}
\begin{samepage}
  \begin{footnotesize}
    \begin{eqnarray}
      y[k] &=& (0.8-0.5\text{exp}(-y[k-1]^2)y[k-1]+u[k-1]- \nonumber\\
             &&(0.3+0.9\text{exp}(-y[k-1]^2)y[k-2] + 0.2u[k-2]+\nonumber\\
           &&0.1u[k-1]u[k-2] + 0.1v[k-1] + 0.3v[k-2] + v[k],\nonumber
    \end{eqnarray}
  \end{footnotesize}
\end{samepage}
\noindent
for which, the values of $u$ are drawn from a
standard Gaussian distribution and held for 5 samples.
And $v$ is a white Gaussian process with standard deviation
$\sigma_v = 0.5$.

\begin{figure}[t]
  \input{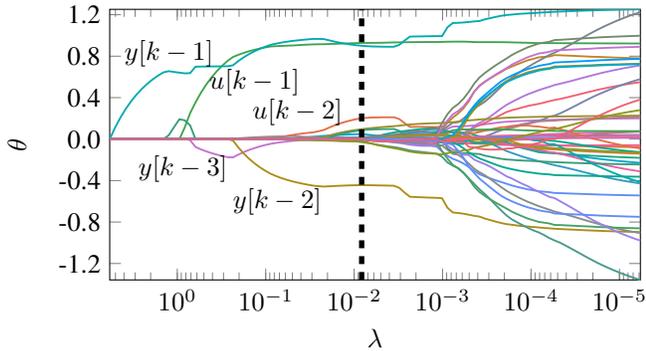}
  \caption{\textbf{(Example 2)} The figure display the estimated parameter
    vectors $\boldsymbol{\theta}$
    as a function of the regularization parameter $\lambda$ (in log-scale).
    A dashed vertical line indicates the value of $\lambda$ which yields
    the best validation results.}
  \label{fig:narmax_example}
\end{figure}

A window of $1000$ samples is used for training
a polynomial model and one with  $500$ samples to validate it.

The following polynomial model will be adjusted to
the training set:
\begin{equation}
  \label{eq:polynomial_model}
  y[k] = \sum_{i}\theta_i \left(y[k-q_i] \right)^{l_i} \left(u[k-t_i] \right)^{r_i}
  \left(e[k-w_i] \right)^{s_i},
\end{equation}
for which the monomials included as regressors are all
possible monomials for which: ${1\le q_i \le n_y}$;
${1 \le t_i \le n_u}$; 
${1 \le w_i \le n_e}$; and, $l_i + r_i + s_i \le n_{\text{degree}}$. In this example,
we have used $n_u=3$, $n_y=3$, $n_e=2$
and $n_{\text{degree}}=2$,
which yields a total number of regressors
$p = 44$.
\begin{table*}
  \centering
  \begin{tabular}{|c|c|c|c|c|c|c|c|c|c|c|c|}
    \multicolumn{3}{c}{\large{a)}} & \multicolumn{9}{c}{$n_y = 3$, $n_u = 3$, $n_e = 0$ (NARX)}\\
    \cline{4-12}
    \multicolumn{3}{c}{} &  \multicolumn{3}{|c|}{$\lambda_{\min} = 10^{-2}\lambda_{\max},~K=100$} & \multicolumn{3}{|c|}{$\lambda_{\min} = 10^{-4}\lambda_{\max},~K=200$} & \multicolumn{3}{|c|}{$\lambda_{\min} = 10^{-6}\lambda_{\max},~K=300$} \\
    \hline
     $n_{\text{degree}}$ & $p$  & $p_e/p$ &  $N = 100$ & $1000$ & $10000$ & $N = 100$ & $1000$ & $10000$ & $N = 100$ & $1000$ & $10000$  \\
    \hline
    2 & 27 & 0.00 & 0.02 & 0.03 & 0.34 & 1.55 & 1.65 & 12.72 & 2.73 & 2.89 & 39.57 \\
    3 & 83 & 0.00 & 0.003 & 0.03 & 0.63 & 1.59 & 1.64 & 20.41 & 9.37 & 28.29 & 488.03 \\
    \hline
    \multicolumn{12}{c}{}\\
    \multicolumn{12}{c}{}\\
    \multicolumn{3}{c}{\large{b)}} & \multicolumn{9}{c}{$n_y = 3$, $n_u = 2$, $n_e = 1$}\\
    \cline{4-12}
    \multicolumn{3}{c}{} &  \multicolumn{3}{|c|}{$\lambda_{\min} = 10^{-2}\lambda_{\max},~K=100$} & \multicolumn{3}{|c|}{$\lambda_{\min} = 10^{-4}\lambda_{\max},~K=200$} & \multicolumn{3}{|c|}{$\lambda_{\min} = 10^{-6}\lambda_{\max},~K=300$} \\
    \hline
    $n_{\text{degree}}$ & $p$  & $p_e/p$ &  $N = 100$ & $1000$ & $10000$ & $N = 100$ & $1000$ & $10000$ & $N = 100$ & $1000$ & $10000$  \\
    \hline
    2 & 27 & 0.22 & 0.02 & 0.03 & 0.36 & 3.07 & 5.89 & 112.98 & 7.33 & 30.31 & 347.54 \\
    3 & 83 & 0.33 & 0.01 & 0.26 & 3.32 & 5.92 & 19.05 & 500.38  & 27.57 & 144.05 & 2724.86 \\
    \hline
    \multicolumn{12}{c}{}\\
    \multicolumn{12}{c}{}\\
    \multicolumn{3}{c}{\large{c)}} & \multicolumn{9}{c}{$n_y = 2$, $n_u = 2$, $n_e = 2$}\\
    \cline{4-12}
    \multicolumn{3}{c}{} &  \multicolumn{3}{|c|}{$\lambda_{\min} = 10^{-2}\lambda_{\max},~K=100$} & \multicolumn{3}{|c|}{$\lambda_{\min} = 10^{-4}\lambda_{\max},~K=200$} & \multicolumn{3}{|c|}{$\lambda_{\min} = 10^{-6}\lambda_{\max},~K=300$} \\
    \hline
    $n_{\text{degree}}$ & $p$  & $p_e/p$ &  $N = 100$ & $1000$ & $10000$ & $N = 100$ & $1000$ & $10000$ & $N = 100$ & $1000$ & $10000$  \\
    \hline
    2 & 27 & 0.44 & 0.01 & 0.02 & 0.20 & 1.10 & 1.87 & 10.41 & 2.75 & 8.31 & 35.51 \\
    3 & 83 & 0.58 & 0.06 & 0.15 & 2.76 & 4.70 & 13.84 & 316.77  & 33.43 & 216.9 & 3415.83 \\
    \hline
    \multicolumn{12}{c}{}\\
    \multicolumn{12}{c}{}\\
    \multicolumn{3}{c}{\large{d)}} & \multicolumn{9}{c}{$n_y = 1$, $n_u = 1$, $n_e = 4$}\\
    
    \cline{4-12}
    \multicolumn{3}{c}{} &  \multicolumn{3}{|c|}{$\lambda_{\min} = 10^{-2}\lambda_{\max},~K=100$} & \multicolumn{3}{|c|}{$\lambda_{\min} = 10^{-4}\lambda_{\max},~K=200$} & \multicolumn{3}{|c|}{$\lambda_{\min} = 10^{-6}\lambda_{\max},~K=300$} \\
    \hline
     $n_{\text{degree}}$ & $p$  & $p_e/p$ &  $N = 100$ & $1000$ & $10000$ & $N = 100$ & $1000$ & $10000$ & $N = 100$ & $1000$ & $10000$  \\
    \hline
    2 & 27 & 0.78 & 0.01 & 0.02 & 0.20 & 0.07 & 0.16 & 1.64 & 0.11 & 0.34 & 3.05 \\
    3 & 83 & 0.88 & 0.01 & 0.23 & 4.27 & 1.98 & 2.78 & 40.64 & 35.85 & 13.23 & 125.65 \\
    \hline
  \end{tabular}
  \caption{Algorithm~\ref{alg:coordinate_optimization} running timings (in seconds)
    for different settings.
    The data used for training was generated as described in Example 2.}
  \label{tab:timings} 
\end{table*}

Again, Algorithm~\ref{alg:coordinate_optimization}
is used to find the Lasso solution for a sequence of
decreasing values of $\lambda$. Figure~\ref{fig:narmax_example}
illustrates the obtained regularization paths. The linear terms are the first to enter the model
and are annotated on the figure.  The value
of $\lambda$ for which the validation error is minimum is indicated with a
dashed vertical line. For this optimal $\lambda$ the mean absolute error
in the validation set is 1.03 and the model includes the terms annotated
in the Figure and 
the regressors $e[k-1]$, $e[k-2]$, ${y[k-1]y[k-2]}$,
$u[k-1]u[k-2]$, $y[k-3]e[k-1]$, ${y[k-2]u[k-2]}$.

\subsection{Timings}
\label{sec:timings}

Table~\ref{tab:timings} shows the algorithm running timings for different
settings. The data used for training is generated as in
Example 2 and the total length of the training dataset is
denoted by $N$. A polynomial as the one described
in (\ref{eq:polynomial_model})
with maximum lags $n_y$, $n_u$ and $n_e$
and order $n_{order}$  is fitted to this data set.
The total number of regressors is denoted
by $p$ and consist of all possible combinations
of monomials within this lag and order constraints.
The fraction of monomials that contains some
error term is denoted by $\tfrac{p_e}{p}$.
Furthermore, different parameters of $\lambda_{\min}$
and $K$ are used in the different experiments.
The stop criteria used is $\|\theta^+ - \theta\|_{\infty} < 10^{-7}$.
All timings were carried out on an Intel
Core i7-4790K 4.00GHz processor.

The more obvious point that can be taken
from Table~\ref{tab:timings} is that under similar
conditions the time grows with both the data set length $N$
and the number of regressors $p$.

The run time also grows if we decrease $\lambda_{\min}$ and
increase the number of points $K$. That is because:
i) the increase on $K$ produces more values of
$\lambda$ to be evaluated;
and,  ii) for smaller values of $\lambda$
the number of non-zero parameters increases, and the
speed up provided by iterating only on the active set
(described in Section~\ref{sec:active-set-conv}) loses its effect.
The importance of effect (ii) can be observed in Table~\ref{tab:timings}
by noticing that under similar conditions
the simultaneous variation of $\lambda_{\min}$ and $K$
often results in a much greater increase
of the running time than what the increment of
the number of points $K$ could account
for.

It follows from the above discussion that
the algorithm computes the first terms to enter the active
set very efficiently due to the sparse structure of the solution.
Hence, a subset of the regressors parameters can usually be efficiently
computed in few inexpensive iterations.

Algorithm~\ref{alg:coordinate_optimization} modifies the original
coordinate descent algorithm by introducing
the step 1-(a), which require the regressor matrix columns to be
updated along the iterations. The fraction $p_e/p$ gives the
number of columns which actually requires to be updated. In
Table~\ref{tab:timings}, it is possible to find entries
that have increasing values of $p_e/p$ for similar
configurations and the run time does not consistently grows
with it. This is a good indicator that the modifications
we introduced in the algorithm are
not critical to the total computation time and that
other aspects, as the correlation between the variables,
may have a much greater influence on
the total running time.

\section{Final Comments}
\label{sec:final-comments}

In this paper we proposed a new pathwise coordinate descent
algorithm for estimating NARMAX models with $L_1$-norm regularization.
To the best of authors' knowledge it is the first algorithm
to consider the inclusion of error terms in the Lasso regression
problem. The time results in \hbox{Section~\ref{sec:timings}} suggests
that the proposed modification does not
reduce the computational efficiency of the original algorithm
and that the computation is especially efficient
when only the more important terms to enter the model are required,
as it is often the case.

Like the extended least squares, the algorithm uses heuristics
that make it very hard to establish mathematical convergence properties.
Nevertheless, the algorithm has converged to meaningful solutions
in all tested situations.

While we have focused on the \textit{Lasso},
the procedure could easily be adapted to \textit{elastic net}
penalties, using a similar reasoning as the one used in~\cite{friedman_regularization_2010}.
The algorithm seems to be very promising and the results presented here suggest it might
prove to be useful in a variety of identification problems.

\section*{Acknowledgments}

This work has been supported by the Brazilian agencies CAPES, CNPq and FAPEMIG.

\bibliographystyle{IEEEtran}
\bibliography{2018-lasso-narmax}

\end{document}